\documentclass[twocolumn,preprintnumbers,amsmath,amssymb]{revtex4}
\usepackage{epsfig}
\usepackage{color}
\usepackage{dsfont}
\usepackage[colorlinks,urlcolor=blue,linkcolor=blue,citecolor=cyan]{hyperref}
\begin{document}
\setlength{\voffset}{1.0cm}
\title{Tricritical curve of massive chiral Gross-Neveu model with isospin}
\author{Michael Thies\footnote{michael.thies@gravity.fau.de}}
\affiliation{Institut f\"ur  Theoretische Physik, Universit\"at Erlangen-N\"urnberg, D-91058, Erlangen, Germany}
\date{\today}

\begin{abstract}
We reconsider the two-flavor version of the massive, chiral Gross-Neveu model in 1+1 dimensions. Its phase diagram as a function of
baryon chemical potential, isospin chemical potential and temperature has previously been explored. We recapitulate the results, adding
the  missing tricritical curves. They can be determined exactly by extending the standard stability analysis, using fourth order almost degenerate
perturbation theory. Results for three different bare masses are presented and discussed.
\end{abstract}

\maketitle

\section{Introduction}
\label{sect1}
The two best known Gross-Neveu (GN) models \cite{L1} are the original one with discrete chiral symmetry ($\psi \to \gamma_5 \psi$) and Lagrangian
\begin{equation}
{\cal L}_{\rm GN} = \bar{\psi} i \partial \!\!\!/ \psi + \frac{g^2}{2} \left(\bar{\psi}\psi \right)^2
\label{1.1}
\end{equation}
and the chiral GN model with U(1) chiral symmetry ($\psi \to e^{i \gamma_5 \alpha}\psi$),
\begin{equation}
{\cal L}_{\rm NJL} = \bar{\psi} i \partial \!\!\!/ \psi + \frac{g^2}{2}\left[ (\bar{\psi}\psi)^2 + ( \bar{\psi} i \gamma_5 \psi )^2 \right].
\label{1.2}
\end{equation}   
We will only be dealing with 1+1 dimensions and the large $N_c$ limit of fermions with a U($N_c$) ``color" symmetry. 
Color indices will be suppressed as usual ($\bar{\psi}\psi=\sum_{k=1}^{N_c} \bar{\psi}_k \psi_k$ etc.).
Extending model (\ref{1.2})
to SU(2) chiral symmetry by introducing isospin, we arrive at the Lagrangian
\begin{equation}
{\cal L}_{\rm isoNJL} = \bar{\psi} i \partial \!\!\!/ \psi + \frac{G^2}{2}\left[ (\bar{\psi}\psi)^2 + ( \bar{\psi} i \gamma_5 \vec{\tau} \psi )^2 \right],
\label{1.3}
\end{equation}
familiar from the standard Nambu--Jona-Lasinio (NJL) model in 3+1 dimensions \cite{L2}.
As already indicated in the subscripts of the Lagrangians, we shall refer to (\ref{1.1}) as GN model, (\ref{1.2}) as NJL model and (\ref{1.3}) as NJL model with isospin (isoNJL). 
Actually, the focus of the present work will be on the massive versions of these models obtained by adding a Dirac mass term (bare mass $m_b$)
\begin{equation}
\delta {\cal L}= - m_b \bar{\psi}\psi.
\label{1.4}
\end{equation}
However, by way of introduction,
we find it appropriate to briefly recall what is known about the phase diagrams of models (\ref{1.1})--(\ref{1.3}) in the chiral limit \cite{L3}.

The starting point for solving the isoNJL model (\ref{1.3}) in the large $N_c$ limit is the Dirac Hartree-Fock (HF) equation
\begin{equation}
\left(-i \gamma_5 \partial_x + \gamma^0 S + i \gamma^1 \vec{\tau} \cdot \vec{P} - \mu - \nu \tau_3   \right)\psi = \omega \psi.
\label{1.5}
\end{equation}
We have introduced a baryon chemical potential $\mu$ and an isospin chemical potential $\nu$.
The scalar and pseudoscalar mean fields $S,\vec{P}$ satisfy the following self-consistency conditions,
\begin{eqnarray}
S  & = & -  G^2 \langle \bar{\psi} \psi \rangle ,
\nonumber  \\
\vec{P} & = & - G^2 \langle \bar{\psi} i \gamma_5 \vec{\tau} \psi \rangle ,
\label{1.6}
\end{eqnarray}
where the brackets denote either ground state or thermal averages.
Let us now assume that the charged pseudoscalar condensate vanishes, $\vec{P}_{\perp}=0$. 
Then the HF Hamiltonian becomes diagonal in isospin space with
\begin{equation}
\left(-i \gamma_5 \partial_x + \gamma^0 S \pm i \gamma^1 P_3 - \mu \mp \nu   \right)\psi = \omega \psi
\label{1.7}
\end{equation}
for isospin up and down, respectively. In each isospin channel, the Dirac HF equation reduces to that of a single-flavor NJL model. For isospin up, the mean field is $\Delta=S-iP_3$ and the chemical potential $\mu+\nu$.
For isospin down, the corresponding parameters are $\Delta=S+iP_3$ and $\mu-\nu$. Although the HF Hamiltonian is diagonal in isospin, the two isospin channels are still coupled through the self-consistency condition, as $P_3$ involves the 
difference between up and down contributions.

\begin{figure}
\begin{center}
\epsfig{file=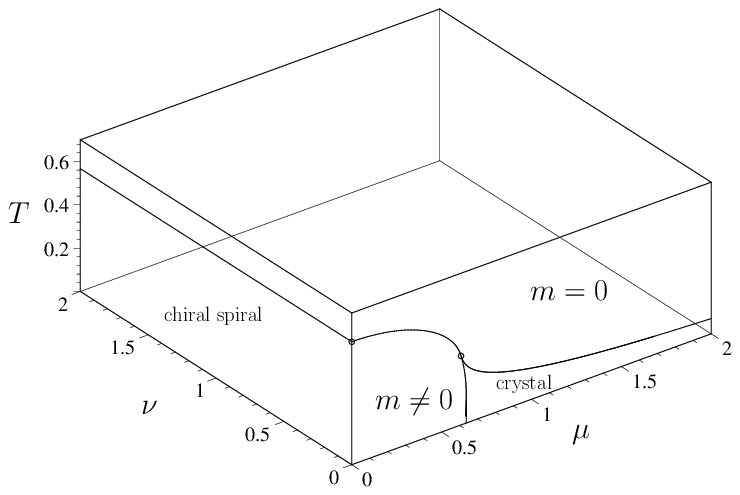,width=8cm,height=7cm,angle=0}
\caption{Boundary phase diagram of massless isoNJL model if one of the chemical potentials vanishes. At $\nu=0$, same as massless GN model. At $\mu=0$, same as 
massless NJL model with chemical potential $\nu$.}
\label{fig1}
\end{center}
\end{figure}

It is now easy to see that the phase diagram of the isoNJL model in ($\mu,\nu,T$) space can be constructed from the known phase diagrams of the 
GN and NJL models in ($\mu,T$) space. To this end, consider first the special cases where one of the chemical potentials vanishes. For $\nu=0$ (pure baryon chemical potential),
the NJL  equations for isospin up and down have the same chemical potential $\mu$ and complex conjugate mean fields $\Delta=S \mp iP_3$. This is only possible 
if $\Delta$ is real ($P_3=0$). Thus the isoNJL phase diagram in the $\nu=0$ plane is identical to the GN phase diagram in the ($\mu,T$) plane, as indeed first noticed 
in the numerical study \cite{L4}. The GN phase diagram in turn is known analytically \cite{L5} and features three phases, a chirally restored one, a homogeneous massive one and a
soliton crystal. For $\mu=0$ on the other hand (pure isospin chemical potential $\nu$), the NJL equations (\ref{1.5}) have opposite chemical potentials $\pm \nu$
and the mean fields are complex conjugates. This is exactly what it takes to solve both equations with the standard NJL solution, so that the phase diagram of the isoNJL model in the $\mu=0$
plane is that of the one-flavor NJL model with chemical potential $\nu$. It is also well known analytically and consists of a chirally restored phase and a soliton crystal of ``chiral spiral" type \cite{L6,L7}.
Thus, the phase diagram of the isoNJL model on the boundaries $\nu=0,\mu=0$ is completely determined by the single-flavor GN and NJL phase diagrams, see Fig~\ref{fig1}.

\begin{figure}
\begin{center}
\epsfig{file=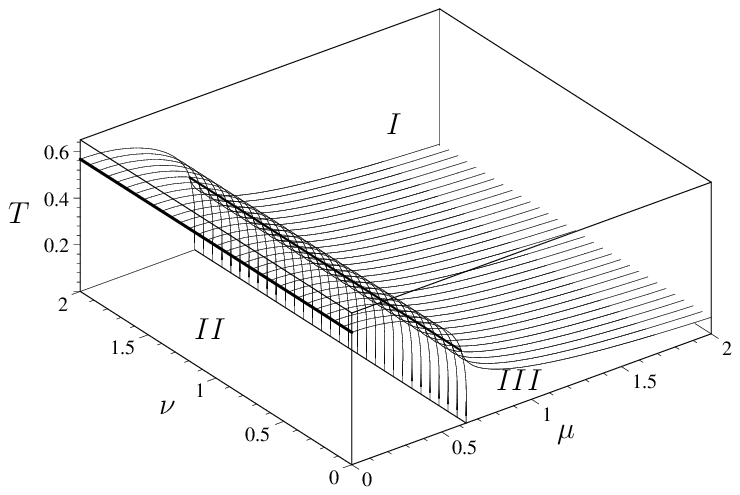,width=8cm,height=7cm,angle=0}
\caption{Phase diagram of massless isoNJL model in the bulk of ($\mu,\nu,T$) space. The phase boundary sheets result from moving the GN phase 
diagram rigidly into the direction of the $\nu$ axis. $I$) chirally restored phase, $II$) chiral spiral with constant radius, $III$) chiral spiral
with radius modulated by GN soliton crystal.}
\label{fig2}
\end{center}
\end{figure}

How can these boundary phase diagrams be continued into the bulk of ($\mu,\nu,T$) space? As noticed in Ref.~\cite{L3}, the same trick used originally to derive the chiral spiral, namely a chiral rotation with linearly $x$-dependent phase, can 
be applied to the isospin case as well. This shows that the phase boundaries of the isoNJL model are independent of $\nu$, just like the phase boundary of the NJL model 
is independent of $\mu$. The full phase diagram can thus be generated by simply translating the GN phase diagram rigidly into the direction of the $\nu$ axis, see Fig.~\ref{fig2}. The resulting mean field
is the product of the GN mean field and the NJL chiral spiral phase factor,
\begin{equation}
S \mp iP_3 = S_{\rm GN}(\mu,T,x) e^{\pm 2i\nu x}.
\label{1.8}
\end{equation}
The structure of a double helix emerges where up- and down contributions have opposite handedness.
This gives rise to three distinct phases of the massless isoNJL model: A chirally restored one ($I$), a double chiral spiral with constant radius ($II$) and a double chiral spiral with $x$-dependent radius, modulated by the 
shape of the GN kink crystal ($III$). One finds that this solution is self-consistent, in spite of the fact that $P_{1,2}=0$ has been assumed from the outset. No better HF solution (with neutral and
charged pion condensates) has been found so far, including the earlier variational calculations \cite{L8,L9}.

This is the status of the phase diagrams of models (\ref{1.1})--(\ref{1.3}) in the chiral limit. The purpose of the present paper is to continue investigating the phase diagram of the massive isoNJL model.
As already known from the one-flavor models, we cannot expect an analytical solution any more, but have to engage in numerical calculations as well. This complicates matters significantly,
but the phase diagram is also expected to be richer than in the chiral limit.

The paper is organized as follows. In Sect.~\ref{sect2} we review what is known about the phase diagram of the massive isoNJL model to date. In Sect.~\ref{sect3} we outline a method
recently proposed to find the exact tricritical point in the massive NJL model. In Sect.~\ref{sect4} we adapt this method to the massive isoNJL model and present detailed
results for the full phase diagram, including for the first time the tricritical curve in ($\mu,\nu,T$) space.

\section{Constructing the phase diagram of the massive ${\bf isoNJL}$ model}
\label{sect2}
\begin{figure}
\begin{center}
\epsfig{file=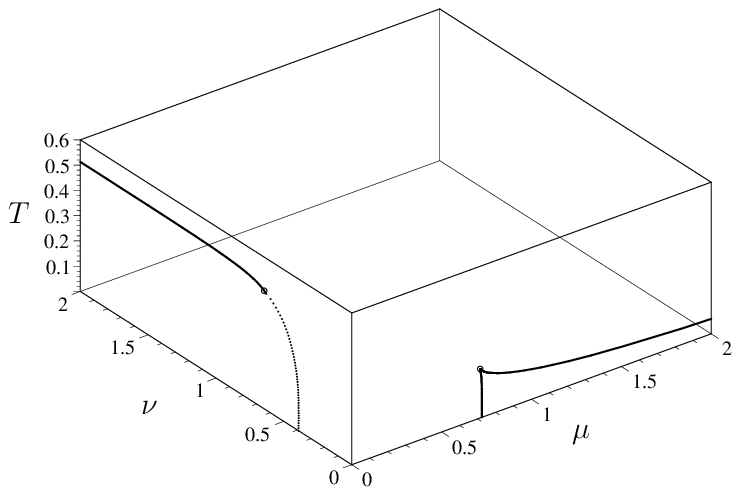,height=7cm,width=8cm,angle=0}
\caption{Boundary phase diagram of massive isoNJL model ($\gamma=0.1$) if one of the chemical potentials vanishes. At $\nu=0$, same as massive GN model. At $\mu=0$, same as 
massive NJL model with chemical potential $\nu$.} 
\label{fig3}
\end{center}
\end{figure}

We now turn to the massive versions of models (\ref{1.1})--(\ref{1.3}) by adding the Dirac mass term (\ref{1.4})
to each Lagrangian. Here the phase diagrams of the GN and NJL model are again known analytically (massive GN \cite{L10}) or at least numerically (massive NJL \cite{L11}). 
Recall that during renormalization, the two bare parameters $m_b$ and $G^2$ are traded for the physical fermion mass in vacuum ($m$) and the so-called 
``confinement parameter" ($\gamma$) according to the gap equation
\begin{equation}
\frac{\pi}{2N_c G^2} = \gamma + \ln \frac{\Lambda}{m}, \quad \gamma= \frac{\pi m_b}{2 N_c G^2m} = \frac{m_b}{m} \ln \frac{\Lambda}{m}.
\label{2.1}
\end{equation} 
This version belongs to the two-flavor isoNJL model (\ref{1.3}). In the case of the one-flavor models (\ref{1.1}), (\ref{1.2}), replace $G^2$ by $g^2/2$. 
We use units such that $m=1$ for any $\gamma$ in the following. 
The Dirac HF equation for the massive isoNJL model is unchanged as compared to (\ref{1.5}), but the self-consistency conditions (\ref{1.6}) now read
\begin{eqnarray}
S - m_b & = & -  G^2 \langle \bar{\psi} \psi \rangle ,
\nonumber  \\
\vec{P} & = & - G^2 \langle \bar{\psi} i \gamma_5 \vec{\tau} \psi \rangle .
\label{2.2}
\end{eqnarray}
We assume once again that the charged components of $\vec{P}$ vanish. Then the reasoning used in the chiral limit goes through
literally, yielding a Hamiltonian diagonal in isospin space and Eq.~(\ref{1.5}) for up- and down quarks. Let us start again by looking at the special cases where one
of the chemical potentials vanishes. The problem then reduces to the single-flavor GN ($\nu=0$) and NJL ($\mu=0$) phase diagrams, now for the massive 
models. Here, only two phases exist, a massive homogeneous one and an inhomogeneous one with spatially periodic order parameter, see Fig.~\ref{fig3}. 
The massless phase is forbidden since chiral symmetry is explicitly broken by the bare mass. 
On the NJL side, the ``horizontal", solid line is a perturbative, 2nd order phase boundary determined via a stability analysis. The ``vertical", dotted curve 
is a first order phase boundary inferred from a numerical HF calculation. The homogeneous solution and a periodic inhomogeneous solution with finite
amplitude coexist along this line. These two curves meet at a tricritical point indicated by a dot. Originally, 
this point had been found by pushing the numerical HF calculation towards the endpoint of the first order line (``bottom up" approach \cite{L12}).
More recently, a new method based on next-to-leading-order (NLO) perturbation theory has been devised to find the exact position of the tricritical point from the 
perturbative side (``top down" approach \cite{L13}), superseding the earlier numerical result. On the GN side, the horizontal part also belongs to a 2nd order
phase transition accessible via a stability analysis. Upon crossing it, the system becomes unstable against the creation of a periodic structure of infinitesimal amplitude.
The vertical part is non-perturbative. Here, the system is unstable against formation of a single baryon. At the cusp, there is a tricritical point and the wave number of the
inhomogeneous phase vanishes. The massive GN results have been obtained in an analytical way \cite{L10}. Let us also mention that the base points of the phase boundaries at $T=0$
are located at the masses of the most strongly bound baryons in both models. Their values are known from independent works, analytically in the GN model \cite{L14} and 
numerically in the NJL model \cite{L15}.

Due to the different character of the phase diagrams on the boundaries $\nu=0$ and $\mu=0$, it is interesting to study how the massive isoNJL model will
manage to interpolate between the two graphs if both chemical potentials are non-vanishing.
Fig.~\ref{fig3} immediately shows that the interpolating phase boundary sheet must depend non-trivially on all three coordinates, in contrast to the 
chiral limit of Fig.~\ref{fig2}. An inspection of Fig.~\ref{fig3} suggests to split the problem into four distinct questions:
How are the horizontal, perturbative phase boundaries connected? What is the curve in the $T=0$ plane, connecting the two base points and separating
homogeneous from inhomogeneous phases at zero temperature? What is the shape of the phase boundary sheet connecting the two vertical, non-perturbative curves?
And, finally, what is the shape of the tricritical curve connecting the two tricritical points? The first three questions have already been answered \cite{L3,L12}. We illustrate the
solution in subsections \ref{sect2a}--\ref{sect2c}, using as example the case $\gamma=0.1$. The last question is the main topic of the present work
and will be covered in Sects.~\ref{sect3} and \ref{sect4}. 

\subsection{Perturbative sheet and stability analysis}
\label{sect2a}
If we set $P_1=P_2=0$, the grand canonical potential of the isoNJL model can be written as a sum over two NJL model expressions,
\begin{eqnarray}
\Psi_{\rm isoNJL} (\mu,\nu,T,S,P_3) & = &  \Psi_{\rm NJL}(\mu+\nu,T,S-iP_3)
\label{2.3} \\
 & +  &  \Psi_{\rm NJL}(\mu-\nu,T,S+iP_3).
\nonumber
\end{eqnarray}
This equation holds provided we use $G^2=g^2/2$ and the same $\gamma$ parameter on both sides. Solving the isoNJL model
is therefore closely related to solving the NJL model. We start with the easiest part of the phase diagram, the perturbative 2nd order phase boundaries
separating the homogeneous from the inhomogeneous phases. The stability analysis consists in the following steps: perturb the massive Dirac Hamiltonian 
by a harmonic potential of the form
\begin{equation}
V=\gamma^0 2 S_1 \cos (2Qx) -i \gamma^1 2P_1 \sin (2Qx).
\label{2.4}
\end{equation}
Evaluate analytically the shift of the single particle energies in 2nd order perturbation theory. Expand the grand canonical potential to leading order in 
the correction. This still has to be minimized with respect to $m,Q,S_1,P_1$. At the phase boundary, $m$ is the same as the fermion mass in the homogeneous phase.
Due to (\ref{2.3}), many formulas can be taken over from the NJL model. The minimizations amount to
setting the Hessian determinant and its derivative with respect to $Q$ equal to 0, 
\begin{equation}
{\rm det} {\cal M} = 0, \quad \partial_Q {\rm det} {\cal M} = 0,
\label{2.5}
\end{equation} 
with ${\cal M}$ the Hessian  matrix
\begin{equation}
{\cal M} = \left( \begin{array}{cc} \partial_{S_1}^2 \Psi  & \partial_{S_1}\partial_{P_1} \Psi \\   \partial_{P_1}\partial_{S_1} \Psi & \partial_{P_1}^2  \Psi \end{array} \right).
\label{2.6}
\end{equation}
In such a leading order (LO) stability analysis, not all of the parameters can be determined. Aside from the critical temperature as a function 
of $\mu,\nu$, one can extract the ratio $R=S_1/P_1$ and the wave number $Q$, but not the overall strength of the perturbation.
$R$ and $Q$ characterize the unstable mode. They are not immediately relevant for the location of the phase boundary, but will
play a role for finding the tricritical point.

\begin{figure}
\begin{center}
\epsfig{file=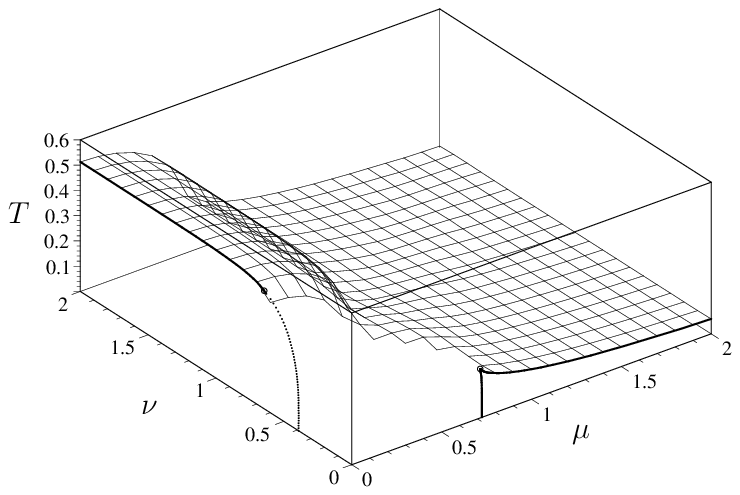,height=7cm,width=8cm,angle=0}
\caption{Adding the perturbative 2nd order sheet to the phase diagram of Fig.~\ref{fig3}. The surface is the result of a LO stability analysis \cite{L3}
and interpolates between the horizontal, perturbative 2nd order phase boundaries of GN and NJL model at $\nu=0$ and $\mu=0$, respectively.} 
\label{fig4}
\end{center}
\end{figure}

Although technically quite simple, such a stability analysis gives already a fairly good impression of the full phase diagram.
It is available for a number of $\gamma$ values \cite{L3}.
In the example at hand ($\gamma=0.1$), the result is shown in Fig.~(\ref{fig4}), smoothly interpolating between GN and NJL perturbative critical curves.
As expected from the boundary phase diagrams at $\mu=0$ and $\nu=0$, this calculation leaves open the details of the homogeneous ``wound" around $\mu=0,\nu=0$.
This will be the subject of the upcoming subsections and sections.

\subsection{Non-perturbative phase boundary in the $T=0$ plane and baryon masses}
\label{sect2b}

\begin{figure}
\begin{center}
\epsfig{file=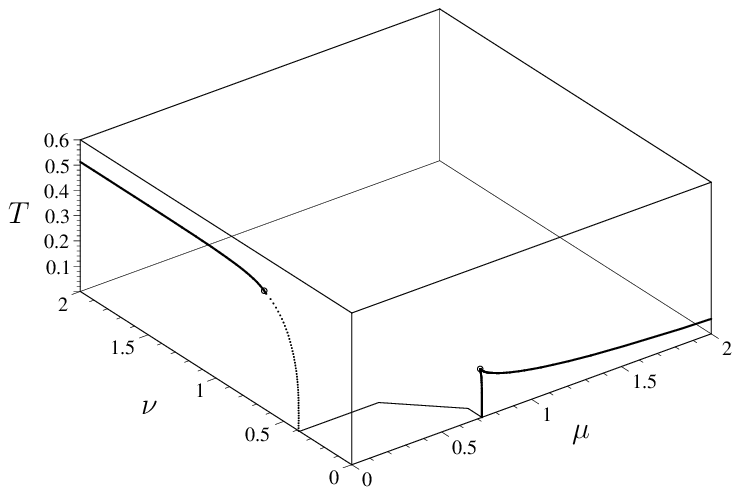,height=7cm,width=8cm,angle=0}
\caption{Adding the non-perturbative phase boundary in the $T=0$ plane to the phase diagram of Fig.~\ref{fig3}. As shown in Ref.~\cite{L12}, it has the shape of the first 
quadrant of an octogon and is fully determined by three baryon masses of the isoNJL model, see main text.} 
\label{fig5}
\end{center}
\end{figure}

The next step is to find the phase boundary at $T=0$, connecting the base points of the GN and NJL phase diagrams. These latter are given by the masses of the most strongly bound baryons
of the two models and are known analytically ($M_{\rm GN}$) or numerically ($M_{\rm NJL}$) in the massive models, as a function of $\gamma$. 
Since these base points are now also part of the isoNJL phase diagram at $T=0$, the two-flavor isoNJL model must possess baryons with the same masses. As shown
in Ref.~\cite{L12}, $M_{\rm GN}$ is the mass of the isoNJL baryon with maximal baryon number and zero isospin, consisting of $N_c$ up quarks and $N_c$ down quarks.
$M_{\rm NJL}$ is the mass of the baryon with zero baryon number and maximal isospin made out of $N_c$ up quarks and $N_c$ down antiquarks.
Many other multifermion bound states with different baryon number and isospin are expected to also play a role along the phase boundary in the $T=0$ plane.
This has been investigated in Ref.~\cite{L12}. Somewhat surprisingly, apart from $M_{\rm GN}$ and $M_{\rm NJL}$, only one other baryon mass enters into the construction
of the phase boundary: the mass $M_{\rm up}$ of the baryon with half maximal baryon number and half maximal isospin, consisting solely of $N_c$ up quarks.
The result for the phase boundary is the first quadrant of an octogon with vertices at the points $(\mu,\nu)=(M_{\rm GN}, 2 M_{\rm up}-M_{\rm GN})$ and 
$(2M_{\rm up}-M_{\rm NJL},M_{\rm NJL})$, see Fig.~\ref{fig5}. It can be made up by intersecting the three lines
\begin{equation}
\nu = M_{\rm NJL}, \quad \mu = M_{\rm GN}, \quad  \mu + \nu = 2 M_{\rm up},
\label{2.7}
\end{equation}
where $M_{\rm NJL} =0.3853$, $M_{\rm GN} = 0.7240$, $M_{\rm up}= 0.4129$ at $\gamma=0.1$. 
Actually, the octogon shape results from constructing the envelope of a whole family of straight lines and seems to be universal
for all $\gamma$ values. In order to construct this family and the envelope, the masses of all possible baryons had to be computed numerically in HF,
even though only three masses are needed eventually.

\subsection{First order sheet at finite $T$ and summary}
\label{sect2c}

\begin{figure}
\begin{center}
\epsfig{file=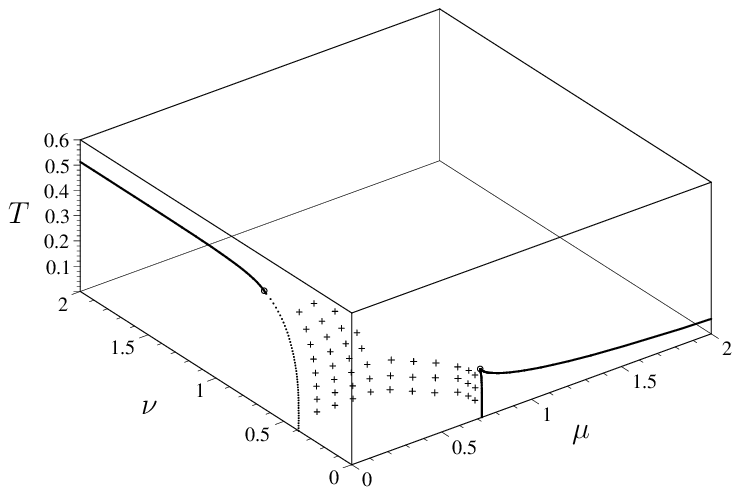,height=7cm,width=8cm,angle=0}
\caption{Adding the first order sheet at $T>0$ (crosses) to the phase diagram of Fig.~\ref{fig3}. It interpolates between the vertical, non-perturbative phase
boundaries of the GN ($\nu=0$) and NJL ($\mu=0$) models. From numerical HF calculation \cite{L12}.} 
\label{fig6}
\end{center}
\end{figure}

The most tedious part of the calculation is the non-perturbative sheet interpolating between the vertical, non-perturbative phase boundaries of GN and NJL models.
It requires a full numerical HF calculation. One has to evaluate the grand canonical potential for the mean fields $S(x),P_3(x)$.
These are assumed to be periodic with wave number $q$ and parameterized in terms of their Fourier components $S_{\ell}, P_{\ell}$. For a given point ($\mu,\nu,T$), one
has to minimize the effective potential with respect to all the $S_{\ell},P_{\ell}$ and $q$. One chooses a trajectory across the anticipated phase boundary
and compares the result with the homogeneous solution. If one finds two curves  intersecting at a point along the trajectory, this point belongs to
the first order phase boundary sheet. The result for $\gamma=0.1$ from Ref.~\cite{L12} is shown in Fig.~\ref{fig6} (crosses). It interpolates between the non-perturbative
curves on the boundary and also matches nicely onto the octogon shape of the $T=0$ phase boundary. The technique used here is identical to the one developed
previously for the massive NJL model, so that we do not go into further details. All points shown were clearly identified as 1st order transitions.

\begin{figure}
\begin{center}
\epsfig{file=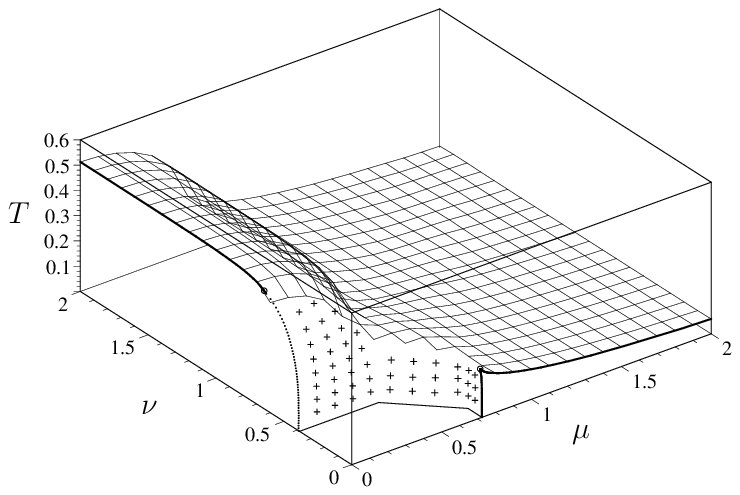,height=7cm,width=8cm,angle=0}
\caption{Present status of the massive isoNJL phase diagram at $\gamma=0.1$, constructed by putting together the ingredients shown in Figs.~\ref{fig3}--\ref{fig6}.
A tricritical curve connecting the tricritical points of the GN and NJL models on the boundaries is still missing. From \cite{L12}.} 
\label{fig7}
\end{center}
\end{figure}

Finally, we put all ingredients discussed so far together in one plot, Fig.~\ref{fig7}. This summarizes the state of the art of the massive isoNJL phase diagram at present,
for the example of $\gamma=0.1$. We find a consistent picture, supporting our assumption that $\vec{P}_{\perp}=0$. All the different
pieces based on independent calculations and a variety of techniques fit together very well, like the pieces of a a jigsaw puzzle. The most glaring deficit is probably the 
fact that we do not know yet how to interpolate between the tricritical points. Hence the line separating the first and second order sheets remains poorly defined. 
As mentioned in Ref.~\cite{L12}, numerical HF calculations in this region were not precise enough, a difficulty first encountered in the massive NJL model.
Recently, a better method has been proposed, tailored to the tricritical points and in principle exact \cite{L13}. This will be reviewed and applied to the case at hand,
the massive isoNJL model, in the next two sections.

\section{Precise determination of the tricritical curve}
\label{sect3}

\begin{figure}
\begin{center}
\epsfig{file=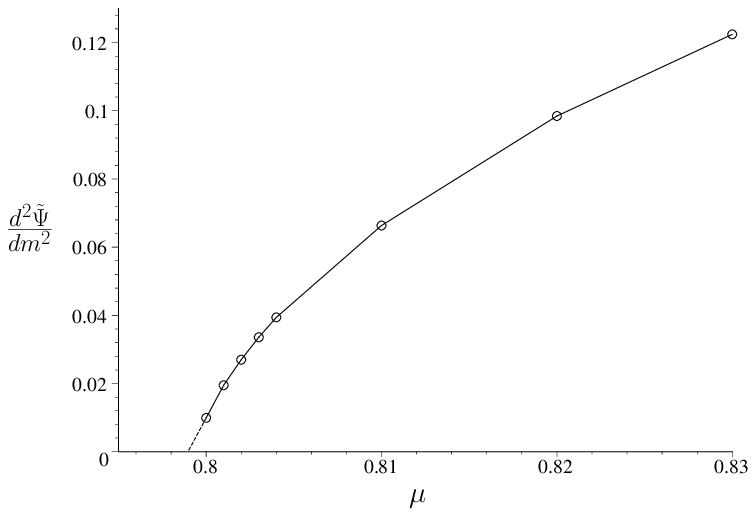,height=6cm,width=8cm,angle=0}
\caption{Proof that the NLO stability analysis developed for the NJL model is also capable of locating the tricritical point of the massive GN model.
The example $\gamma=0.3$ is shown.
The 2nd derivative of the effective potential could be followed as a function of $\mu$ down to $\mu=0.8$ (circles). Extrapolation to 0 yields the value
$\mu=0.799$, in excellent agreement with the known tricritical point.} 
\label{fig8}
\end{center}
\end{figure}

Recently, a novel method of locating the tricritical point has been devised and tested successfully in the massive NJL model \cite{L13}. The basic idea
is to start from the stability analysis,  but pushing perturbation theory to NLO (4th order in $S_1,P_1$). In the case of
the single-flavor NJL model, the procedure may be summarized as follows:
\begin{enumerate}
\item Find a point on the perturbative sheet close to where the tricritical point is expected via a LO stability analysis. Find the coordinates ($\mu,\nu,T$)
together with the ratio $R=S_1/P_1$ and the wave number $Q$ of the unstable mode.
\item Do a 4th order perturbative evaluation, first of the HF single particle energies and then of the thermodynamic potential, keeping $R,Q$ 
fixed at the LO values. The remaining parameters are $m,P_1$. 
\item Choose 3 masses $m_0, m_0\pm \Delta m$ with  $m_0$ the mass of the homogeneous solution at point ($\mu,\nu,T$) on
the perturbative sheet and $\Delta m \ll m_0$. Minimize the grand canonical potential for each mass value with respect to $P_1$. The resulting effective potential
is a function of $m$ only,
\begin{equation}
\widetilde{\Psi}(m) = \min_{P_1} \Psi(m,P_1).
\label{3.1}
\end{equation}
\item The criterion for the tricritical point is the vanishing of the 2nd derivative of the effective potential, $\partial_m^2 \left. \widetilde{\Psi}\right|_{m_0}$  or, in discretized form,
\begin{equation}
\frac{\widetilde{\Psi}(m_0+\Delta m) -2 \widetilde{\Psi}(m_0) + \widetilde{\Psi}(m_0-\Delta m)}{\Delta m^2} = 0.
\label{3.2}
\end{equation} 
\end{enumerate}

The main technical difficulty which had to be overcome is the fact that perturbation theory breaks down near the spectral gaps of a periodic
potential. This becomes more serious in higher order PT where the usual LO ``almost degenerate perturbation theory" (ADPT) cannot be
applied any more.
The way out is to go via an effective Hamiltonian in the space of strongly mixed states and diagonalize it exactly, using a formalism from
many-body perturbation theory due to Lindgren \cite{L16}. The method has been shown to work very well for the massive NJL model and is 
supposedly exact, just like the LO stability analysis for the perturbative sheet.

When applying this method to the isoNJL model, all we have to do is use Eq.~(\ref{2.3}) and reduce the technicalities to those of the NJL model.
One question which then arises is:
Can we be sure that the criterion developed for the massive NJL model also applies to the isoNJL model? The nature of the two tricritical points
on the boundaries $\nu=0$ (GN) and $\mu=0$ (NJL) seems to be very different at first glance.
In the NJL model, the two phase boundaries adjacent to the tricritical point meet tangentially under $0^{\circ}$, whereas the corresponding angle is $180^{\circ}$
at the cusp of the GN model. On the other hand, since we have the representation (\ref{2.3}) of the isoNJL model as a sum of two NJL models, it is plausible
that the methods for finding the tricritical points in both models are closely related. As an additional check, we have applied the method of Ref.~\cite{L13} to the 
GN model. Since $P_1=0$, there are minor changes in the stability analysis. The unstable mode
is only characterized by $Q$, there is no ratio $R$. In the 4th order perturbative calculation, one can use the machinery developed in Ref.~\cite{L13} 
by simply setting $P_1=0$. We have evaluated the 2nd derivative (\ref{3.2}) for several $\mu$ values near the tricritical point. The result is shown in Fig.~\ref{fig8} for $\gamma=0.3$. 
We have not quite succeeded in reaching the point where the 2nd derivative vanishes. However, as the figure shows, we can come very close to it. A
linear extrapolation from the two last points then yields the value $\mu=0.799$, in excellent agreement with the known GN tricritical point at the same $\gamma$ 
($\mu= 0.7986$). The same kind of agreement was reached at other values of $\gamma$.  Apparently, the tricritical point in the GN model lies right at the edge of the 
perturbative sheet, but this is not expected for any other point on the tricritical curve. Thus we are confident that the perturbative method of locating the tricritical point can also 
be trusted in the isoNJL model. The results will be presented in the following section.

\section{Results and discussion}
\label{sect4}

\begin{figure}
\begin{center}
\epsfig{file=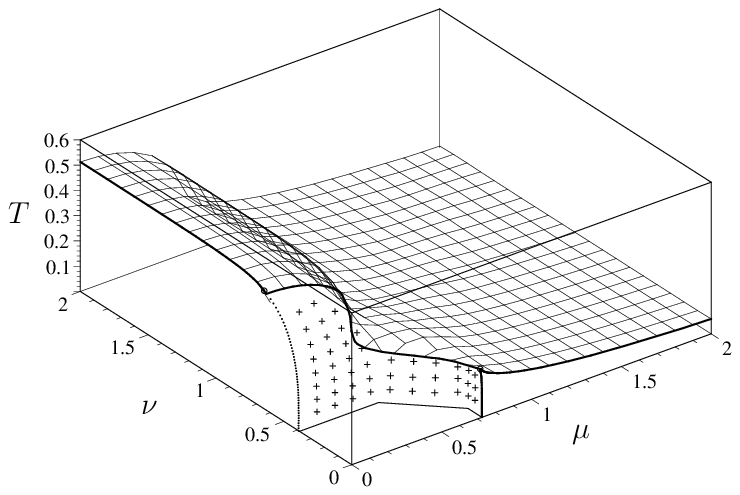,height=7cm,width=8cm,angle=0}
\caption{Final phase diagram for massive isoNJL model at $\gamma=0.1$. The tricritical curve has been added to the previous phase diagram Fig.~\ref{fig7},
now delimiting more clearly second order and first order sheets of the phase boundary between homogeneous and inhomogeneous phases.} 
\label{fig9}
\end{center}
\end{figure}

\begin{figure}
\begin{center}
\epsfig{file=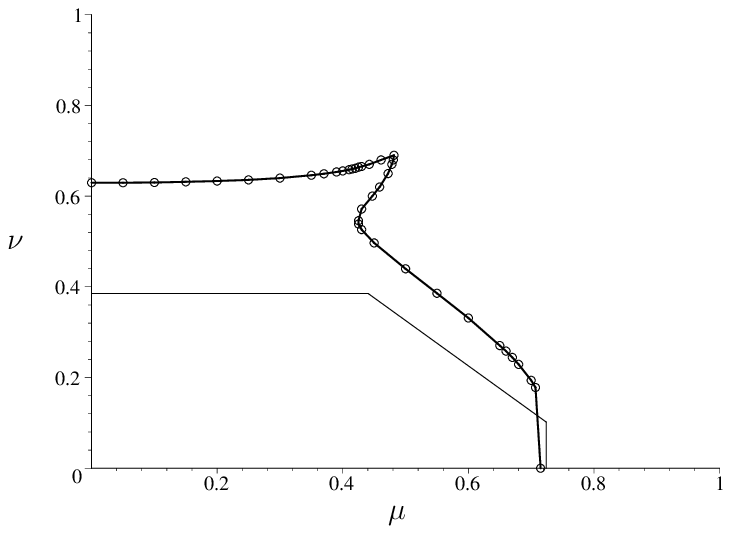,height=6cm,width=8cm,angle=0}
\caption{Projection of tricritical curve onto the $T=0$ plane, for $\gamma=0.1$. The circles are the points actually computed, the fat curve is just an interpolation
which has been used for Fig.~\ref{fig9}. The thin polygon is the phase boundary at $T=0$ and $\gamma=0.1$ discussed in Sect.~\ref{sect2b}. Relevant baryon masses:
$M_{\rm GN}=0.7240, M_{\rm NJL}= 0.3853,M_{\rm up} =0.4129$.}
\label{fig10}
\end{center}
\end{figure}

\begin{figure}
\begin{center}
\epsfig{file=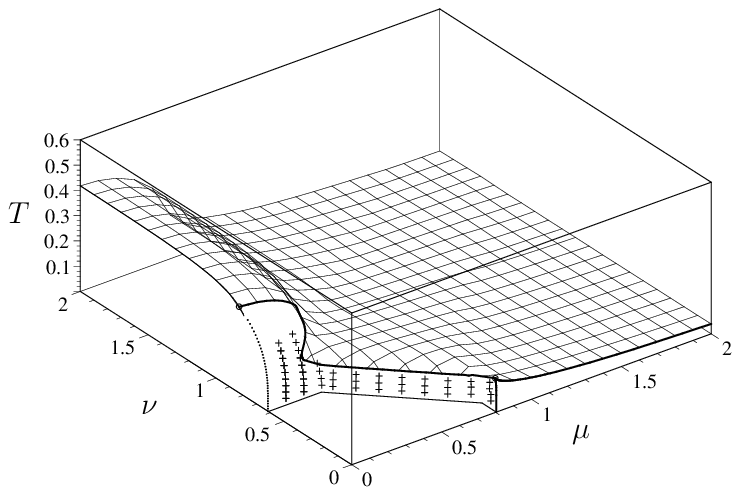,height=7cm,width=8cm,angle=0}
\caption{Same as Fig.~\ref{fig9}, but for $\gamma=0.3$} 
\label{fig11}
\end{center}
\end{figure}

\begin{figure}
\begin{center}
\epsfig{file=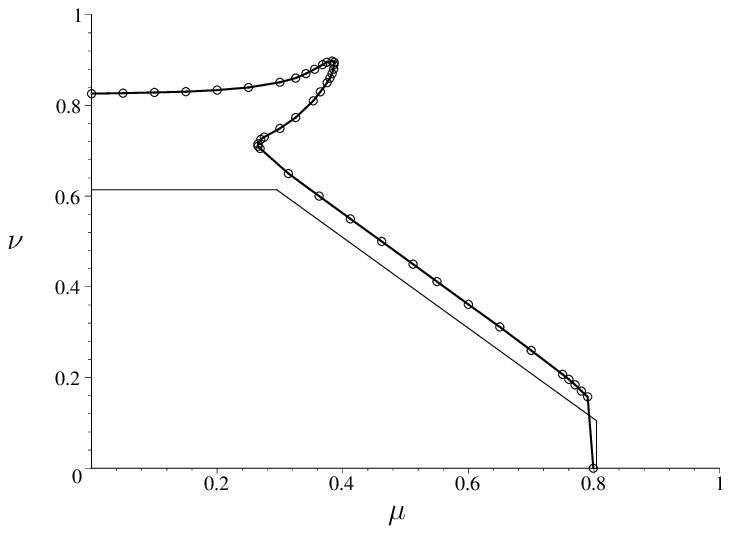,height=6cm,width=8cm,angle=0}
\caption{Same as Fig.~\ref{fig10}, but for $\gamma=0.3$. Relevant baryon masses: $M_{\rm GN}=0.8041, M_{\rm NJL}=0.6142,M_{\rm up} =0.4546$.} 
\label{fig12}
\end{center}
\end{figure}

We have computed the tricritical curve still missing in the phase diagram Fig.~\ref{fig7} for the massive isoNJL at $\gamma=0.1$. The method
has been sketched above and explained in more detail in Ref.~\cite{L13}, where it was applied to the massive NJL model. The main complication of the two-flavor
model is the fact that we now have to determine a full curve of unknown shape and location, except for the endpoints. This means that we have to repeat the
NLO perturbative calculation many times, each calculation being basically the same as for the one-flavor NJL model. Our result for $\gamma=0.1$ is shown in Fig.~\ref{fig9}. The tricritical
curve thus obtained is fully compatible with the other building blocks of the phase diagram and enables us to delimit the inhomogeneous region of the phase 
diagram in a quantitative way. 

\begin{figure}
\begin{center}
\epsfig{file=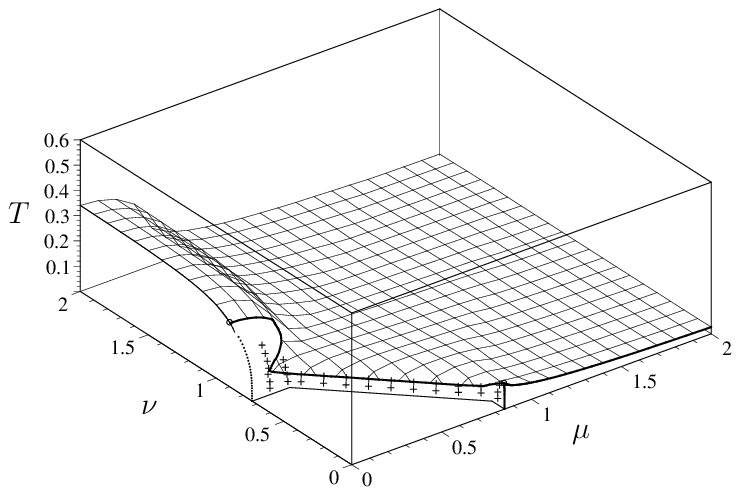,height=7cm,width=8cm,angle=0}
\caption{Same as Fig.~\ref{fig9}, but for $\gamma=0.5$} 
\label{fig13}
\end{center}
\end{figure}

\begin{figure}
\begin{center}
\epsfig{file=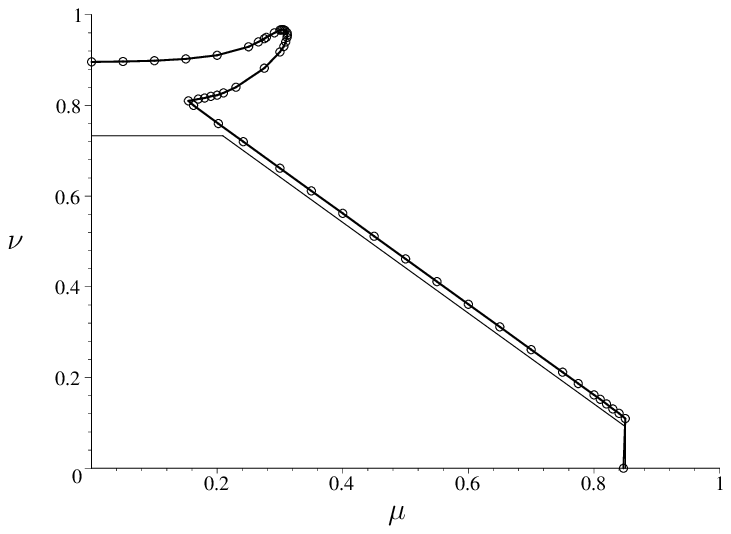,height=6cm,width=8cm,angle=0}
\caption{Same as Fig.~\ref{fig12}, but for $\gamma=0.5$. Relevant baryon masses: $M_{\rm GN}=0.8501, M_{\rm NJL}=0.7334 ,M_{\rm up} =0.4710$.} 
\label{fig14}
\end{center}
\end{figure}

In order to exhibit the shape of the tricritical curve in greater detail, we propose to look at its projection onto the $T=0$ plane.
The result is shown in Fig.~\ref{fig10} for $\gamma=0.1$. We have marked the points which have actually been 
calculated by circles. The fat solid line is an interpolating curve and corresponds to the tricritical curve shown in the 3d Fig.~\ref{fig9}.
Due to the complicated shape of the tricritical curve, a fairly large number of points was needed. 
We have also included the $T=0$ phase boundary into Fig.~\ref{fig10}, the thin polygon. The tricritical curve and the base curve are obviously correlated. 
At $T=0$, we know that the instability along the three straight line segments is determined by three different baryons. Correspondingly, we shall refer to the three segments
as GN (near $\nu=0$), up (tilted) and NJL (near $\mu=0$) parts. A similar   
division can be made for the first order sheet and the tricritical line. The GN and NJL parts are still strongly influenced by the boundary phase diagrams. 
In these regions, the temperature drops to a common level. The up region seems to be the one with the simplest features. The projected tricritical curve follows the slope of
the line $\mu+\nu={\rm const.}$, the first order sheet is steep and essentially planar and the temperature is slowly varying.
This is perhaps the most interesting part, since it has no correspondence in 
the one-flavor phase diagrams.  As speculated in Ref.~\cite{L13}, it is quite likely that there are additional 
phase boundaries inside the crystal phase, starting from the vertices of the $T=0$ octogon and going inside the crystal region. If this is true, the question arises
what happens if we go up in temperature. If the internal phase boundary would persist all the way through the inhomogeneous phase, there should be phase boundaries
in addition to the tricritical curve discussed so far. We have searched for this with our method of finding tricritical points, but without success. This indicates that if
such internal phase boundaries really exist at $T=0$, they disappear at some temperature below the perturbative sheet.

\begin{figure}
\begin{center}
\epsfig{file=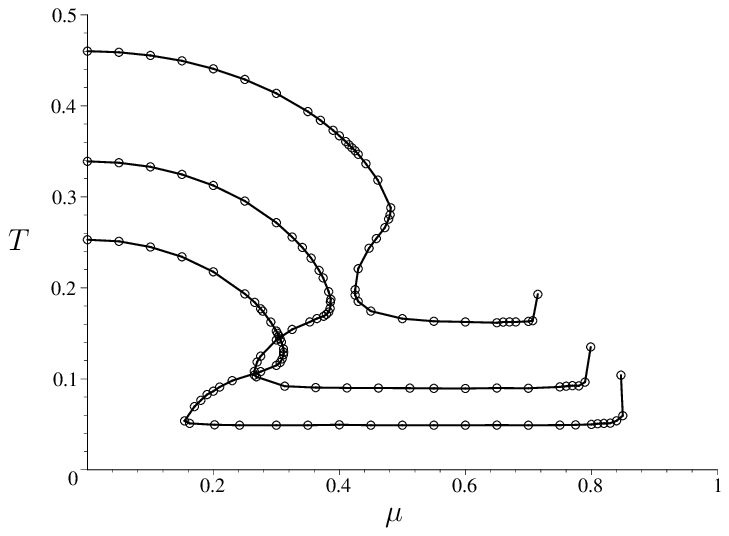,height=6cm,width=8cm,angle=0}
\caption{Projection of tricritical curves onto the $\nu=0$ plane. From top to bottom: $\gamma=0.1,0.3,0.5$.} 
\label{fig15}
\end{center}
\end{figure}

\begin{figure}
\begin{center}
\epsfig{file=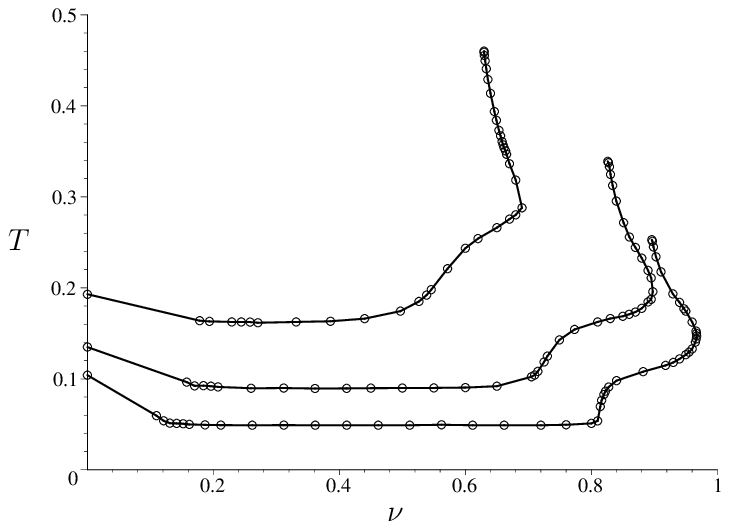,height=6cm,width=8cm,angle=0}
\caption{Projection of tricritical curves onto the $\mu=0$ plane. From top to bottom: $\gamma=0.1,0.3,0.5$.} 
\label{fig16}
\end{center}
\end{figure}

In Ref.~\cite{L13}, the non-perturbative sheet of the massive isoNJL phase diagram has been evaluated at three values of $\gamma$, 0.1, 0.2, and 0.3.
Therefore we have also repeated the calculation of the tricritical curve for $\gamma=0.2$ and $\gamma=0.3$ to see the evolution with increasing bare
fermion mass. The 3d plots are shown in Figs.~\ref{fig11} and \ref{fig13}, along with the projections onto the $T=0$ plane and the $T=0$ phase boundaries 
in Figs.~\ref{fig12} and \ref{fig14}.
Qualitatively, the three phase diagrams and tricritical curves at the three different $\gamma$ values look very similar. As is particularly striking in the projections,
the up part becomes more dominant with increasing $\gamma$ and the simpler features of that part of the phase boundary show up more clearly.
Thus, at the highest
$\gamma$, the $T=0$ line and the projection of the tricritical curve are almost indistinguishable and account for a large part of the phase boundary, see Fig.~\ref{fig14}. 
The strong variation of the projected tricritical curve near the largest $\nu$ values (the ``nose") is a reflection of the part of the curve which rises
steeply in temperature, as can be inferred from the 3d plot, Fig.~\ref{fig13}.

As mentioned above, the up parts of the tricritical curves have essentially constant temperature, and at the same time the lowest one along the whole tricritical curve.
In order to exhibit this
effect more clearly, we have also plotted projections of the three tricritical curves onto the ($\mu,T$) and ($\nu,T$) planes, see Figs.~\ref{fig15} and \ref{fig16}.
Common to all bare masses is the observation that the temperature decreases in the GN section, stays constant through the up section and increases more
strongly in the NJL section. This reinforces the impression that the physics in the up section should be most easily accessible, perhaps by using some heavy quark 
approximation for large $\gamma$.


\begin{thebibliography}{99}
\bibitem{L1}
D. J. Gross and A. Neveu, Phys. Rev. D {\bf 10}, 3235 (1974).
\bibitem{L2}
Y. Nambu and G. Jona-Lasinio, Phys. Rev.  {\bf 124}, 246 (1961).
\bibitem{L3}
M. Thies, Phys. Rev. D {\bf 101}, 014010 (2020).
\bibitem{L4}
A. Heinz, F. Giacosa, M. Wagner, D. H. Rischke, Phys. Rev. D {\bf 93}, 014007 (2016).
\bibitem{L5}
O. Schnetz, M. Thies, and K. Urlichs, Ann. of Phys. {\bf 314}, 425 (2004).
\bibitem{L6}
V. Sch\"on and M. Thies, {\it At the Frontier of Particle Physics: Handbook of QCD, Boris Ioffe Festschrift},
vol. 3, ed. M. Shifman (Singapore: World Scientific), ch. 33, p. 1945 (2001).
\bibitem{L7}
G. Basar, G. V. Dunne, and M. Thies, Phys. Rev. D {\bf 79}, 105012 (2009).
\bibitem{L8}
T. G. Khunjua, K. G. Klimenko, R. N. Zhokhov, V. C. Zhukovsky, Phys. Rev. D {\bf 95}, 105010 (2017).
\bibitem{L9}
T. G. Khunjua, K. G. Klimenko, and R. N. Zhokhov, Phys. Rev. D {\bf 100}, 034009 (2019).
\bibitem{L10}
O. Schnetz, M. Thies, and K. Urlichs, Ann. of Phys. {\bf 321}, 2604 (2006).
\bibitem{L11}
C. Boehmer, U. Fritsch, S. Kraus, M. Thies, Phys. Rev. D {\bf 78}, 065043 (2008).
\bibitem{L12}
M. Thies, Phys. Rev. D {\bf 101}, 074013 (2020).
\bibitem{L13}
M. Thies, Phys. Rev. D {\bf 105}, 116003 (2022).
\bibitem{L14}
M. Thies and K. Urlichs, Phys. Rev. D {\bf 71}, 105008 (2005).
\bibitem{L15}
C. Boehmer, F. Karbstein, and M. Thies, Phys. Rev. D {\bf 77}, 125031 (2008).
\bibitem{L16}
I. Lindgren, J. Phys. B: Atom. Molec. Phys. {\bf 7}, 2441 (1974).
\end{thebibliography}
\end{document}